\documentstyle[12pt,aaspp4]{article}
\def\gs{\mathrel{\raise0.35ex\hbox{$\scriptstyle >$}\kern-0.6em 
\lower0.40ex\hbox{{$\scriptstyle \sim$}}}}

\def\ls{\mathrel{\raise0.35ex\hbox{$\scriptstyle <$}\kern-0.6em 
\lower0.40ex\hbox{{$\scriptstyle \sim$}}}}
\slugcomment{submitted to Publications of the Astronomical Society of
the Pacific}

\lefthead{Wright \& Brainerd}
\righthead{Anisotropic Galaxy--Galaxy Lensing}

\begin{document}

\title{Anisotropic Galaxy--Galaxy Lensing}

\author{Tereasa G. Brainerd \& Candace Oaxaca Wright}
\affil{Boston University, Department of Astronomy, Boston, MA 02215}

\begin{abstract}
We investigate the weak lensing shear due to dark matter galaxy halos whose
mass distributions, 
as projected on the sky, are nearly elliptical.  The shear pattern due to
these halos is anisotropic about the lens centers
and we quantify the level of anisotropy by comparing
the mean shear experienced by sources located closest to the major axes of the
lenses, $\left< \gamma \right>_{\rm major}$,
to that experienced by sources located closest to the minor axes,
$\left< \gamma \right>_{\rm minor}$.
We demonstrate 
that the degree of anisotropy is independent of angular scale and show
that in the case of substantially flattened halos
($\epsilon = 0.7$), the value of $\left< \gamma \right>_{\rm minor}$ is
of order 40\% of the value of  
$\left< \gamma \right>_{\rm major}$ 
when all sources within $\pm 45^\circ$ of the axial direction vectors of
the lenses are included in the calculation.
In the case of halos that are flattened at more realistic
level ($\epsilon = 0.3$), the value of $\left< \gamma \right>_{\rm minor}$
is of order 75\% of the value of $\left< \gamma \right>_{\rm major}$.
We compute the degree to which the anisotropy in the lensing signal
is degraded due to a noisy determination of the position angles of the
lens galaxies and find that provided the typical 1-$\sigma$
error on the orientation
of the lenses is less than
$15^\circ$, more than 90\% of the true lensing signal
will be recovered in the mean.
We discuss our results in the context 
of detecting anisotropic galaxy--galaxy lensing in large, ground--based
data sets and conclude that a modest net flattening of 
dark matter halos should be detectable at a statistically significant
level.  The forthcoming Sloan Digital Sky Survey (SDSS) data will
necessarily provide a very useful data set for this analysis, but 
a detection of
anisotropic galaxy--galaxy lensing is not dependent upon
the very large sky coverage of the SDSS.  Rather, we argue that
a significant
detection of this effect can also be obtained from an imaging survey
that is of order two magnitudes fainter than SDSS
and which covers only a relatively small area of sky, 
of order one square degree.

\end{abstract}



\section{Introduction}

The observed flatness of the rotation
curves of spiral galaxies provides convincing evidence that galaxies
reside within massive dark matter halos (see, e.g., Fich \& Tremaine 1991
and references therein); however, there is as yet no strong constraint
on either
the average radial extent or the typical shapes of these apparently pervasive
structures.  Since
it is generally difficult to
find dynamical tracers of the halo potential
at very large radius, direct observational constraints
on the maximum extent of dark matter halos 
are relatively scarce.  Zaritsky \& White (1994) and
Zaritsky et al.\ (1997) have, however, investigated the dynamics of genuine
satellites of bright field spirals and find that their halos extend to
radii beyond $100h^{-1}$~kpc.

If the dark matter halos of bright field galaxies have radial extents that
are, indeed, as
large as suggested by the above investigations and
the characteristic depths of the potential wells of the halos correspond to
velocity dispersions of order 150 km/s, then weak but detectable gravitational
lensing by the dark halos should occur.  
That is, systematically throughout the universe, the halos of
foreground field galaxies should act as gravitational lenses for background
field galaxies, resulting in a slight preference for the images of distant
galaxies to be oriented tangentially with respect to the locations on the 
sky of galaxies which are physically closer to the observer.  
This effect is known as galaxy-galaxy lensing and
its existence has recently been confirmed by a number of independent
investigations
(e.g., Brainerd, Blandford \& Smail 1996, hereafter BBS; 
Dell'Antonio \& Tyson 1996;
Griffiths et al.\ 1996; Hudson et al.\ 1998; Ebbels 1998; Natarajan et al.\
1998; Fischer et al.\ 2000).  None of these studies has provided a
strong constraint on the maximum physical extent of the halos of field
galaxies, but all have yielded reasonably consistent measurements of
the halo velocity dispersion, on the order of 150 km/s for
an $L^\ast$ galaxy.

Galaxy--galaxy lensing has the potential to be a powerful
probe of the gravitational potential of dark halos at very large
radii.  It has an advantage over dynamical methods in that it can be
applied to all galaxies (in particular those without genuine companions) over
large physical scales, but has the disadvantage that the signal is so small
that it can only be detected in a statistical sense from a large
ensemble of lenses and sources.  That is, while the signal is too weak
to be detected from a single lens galaxy and, hence, cannot be used
to place constraints on 
the potential of any one individual dark matter halo, it can be used to 
probe the mean potential of the halo population as a whole.

Because of the ability of the singular isothermal sphere model to reproduce
the observed flatness of the rotation curves of the disks of spiral galaxies,
it is often assumed {\it a priori} that dark halos 
are roughly spherical.
However, there now
exist a number of direct observations which suggest that dark
halos may be substantially flattened.  The evidence
is diverse, consisting of studies of the kinematics of polar ring
galaxies, the geometry of X-ray isophotes, the flaring of HI gas in
spirals, and the evolution of gaseous warps, as well as the kinematics
of Population II stars in our own Galaxy.  In particular, studies of
disk systems
that probe distances of order 15 kpc from the galactic planes
suggest that the shapes of the dark halos are significantly flattened
and can be characterized by $c/a = 0.5 \pm 0.2$ (see, e.g., the
comprehensive review by Sackett 1999
and references therein).
Here $c/a$ is the ratio of the shortest to longest axis in a principal
moment analysis of the mass density of the halo.
More recently, Kochanek et al.\ (2000) have performed a detailed 
analysis of the shapes of the Einstein rings of three different lensed
quasar host galaxies and conclude that the ellipticities of the
projected mass distributions
of the lens galaxies that give rise to these images
are quite large indeed (axis ratios on the order of 0.6 to 0.7).

All of the
published  attempts to use observations of
galaxy--galaxy lensing to constrain the
characteristic physical parameters of dark matter halos have
assumed the halos to be spherically symmetric.  
The lensing signal has been detected via a circular average of the signal
about the lens center and is then interpreted via halo models which are
spherically symmetric.  However, if the halos are flattened at the
level suggested by the above investigations, their projected surface
mass densities will deviate from circular symmetry and, as result,
the gravitational lensing pattern will not be circularly--symmetric
about the lens center.  Instead, the signal will be mildly anisotropic,
with the shear being strongest along direction vectors that coincide
with the major axis of the mass distribution and weakest along direction
vectors that coincide with the minor axis of the mass distribution.

There are as yet no direct observational constraints on the mean
flattening of the population of dark matter galaxy halos as a whole.
This is due to the fact that the above studies which suggest substantial
halo flattening rely on a handful of specific galaxies for which
the particular
analysis technique (e.g., dynamics, hydrodynamics, or strong lensing)
may be applied.  None of these techniques can be applied in general to
the entire population of galaxies, but systematic galaxy--galaxy lensing
can be applied to all galaxies for which high-quality
imaging data is available.  Despite the fact that it is only the mean
2-D shape that may be directly recovered with weak lensing (not the
full 3-D shape of the halos), a strong constraint on any net flattening
of galaxy halos in 2-D will be useful from the standpoint of understanding
the details of
galaxy formation (i.e., the interaction of baryons with the dark matter
and the transfer of angular momentum to the halo as a result)
and constraining the nature of the dark matter itself.  In 
particular, dissipationless cold dark matter models routinely lead
to the formation of galaxy halos with projected ellipticities of
order 0.3 (e.g.,  Dubinski \& Carlberg 1991; Warren et al.\ 1992), while
models of strongly self-interacting cold dark matter give rise to halos
that are nearly spherical within the virial radius, resulting in a 
small projected ellipticity (e.g., Moore et al.\ 2000).

Here we investigate weak lensing due to halos with elliptical
mass distributions and we compare the shear experienced by sources closest
to the major axes of the lenses to that experienced by sources closest
to the minor axes.
In \S2 we describe the surface mass density profile that we
have adopted, we
evaluate the shear as function of location on the sky for lenses with
mass ellipticities in the range $0.1 \le \epsilon \le 0.7$, and we compute
the level of anisotropy in the galaxy--galaxy lensing signal for sources
located within $\pm 45^\circ$ and $\pm 20^\circ$ of the axial direction
vectors of the lenses.  In
\S3 we discuss the prospects for detecting flattened dark halos via
an anisotropy in the
galaxy--galaxy lensing signal, including a consideration of some of
the potential sources of noise that will be encountered in a realistic
data set. A brief summary of our results is presented in \S4.

\section{Anisotropic Weak Shear Due to an Elliptical Lens}

We adopt a surface mass density for our lenses of the form
\begin{equation}
\kappa(\rho) = \frac{\kappa_0}{\left( 1 + \rho^2/x_c^2 \right)^{1/2}}
\end{equation}
(e.g., Schneider \& Weiss 1991, hereafter SW91), 
which for circularly--symmetric lenses
corresponds to an isothermal sphere with a core radius of $x_c$.  Here
$\kappa(\rho)$ is the surface mass density  of the lens in 
units of the critical surface mass density ($\Sigma_c$),
$\kappa_0$ is the central surface mass density,
$\rho$ is a generalized elliptical radius 
($\rho^2 = x_1^2 + f^2 x_2^2$, where $f = a/b$), and the ellipticity of the
mass is $\epsilon = 1 - b/a$.  That is, equation (1) is an expression 
for the convergence of the lens and all redshift information in
the problem is contained within $\Sigma_c$:
\begin{equation}
\Sigma_c = \frac{c^2}{4\pi G} \frac{D_s}{D_d D_{ds}} .
\end{equation}
Here $D_s$ is the angular diameter distance between the observer and
the source, $D_d$ is the angular diameter distance between the observer
and the lens (the ``deflector''), and $D_{ds}$ is the angular diameter
distance between the lens and the source.

SW91 have shown that the deflection angle, 
$\vec{\alpha} = (\alpha_1, \alpha_2)$, due to
this mass distribution can be computed analytically and they have derived
recursion relations for its computation (equations (A13) through
(A19) in their paper).  We have used these recursion relations 
to compute the deflection angles due to our elliptical masses where the
ellipticities are in  the
range of $0.1 \le \epsilon \le 0.7$. 
Also, since the presence of a core radius does not have a significant
effect on the weak lensing regime in which
systematic galaxy--galaxy lensing occurs, we restrict our analysis below to 
the case of lenses with 
negligible core radii ($x_c = 0.005$ arcsec). 

For each lens ellipticity, we have computed the shear due to the lens, 
$\vec{\gamma}(\vec{\theta}) = (\gamma_1(\vec{\theta}),\gamma_2(\vec{\theta}))$,
on regular grids that were centered on the lens centers.  This was done via a 
straightforward differencing technique in which a regular grid of 
light rays
was traced through each of the lenses and the net deflection of each light
ray was computed from the recursion relations for 
$\vec{\alpha}$ contained in SW91.  
If $\vec{\beta}$ is 
the location of a given light ray
on the grid prior to lensing, and $\vec{\theta}$ is the location of the light
ray after having been deflected, then the components of the shear at the
location $\vec{\theta}$ are simply:
\begin{equation}
\gamma_1(\vec{\theta}) = - \frac{1}{2} \left( \frac{\partial
\beta_{\rm x_1}}
{\partial \theta_{\rm x_1}} - \frac{\partial \beta_{\rm x_2}}
{\partial \theta_{\rm x_2}} \right) ,
\end{equation}
\begin{equation}
\gamma_2(\vec{\theta})= - \frac{1}{2} \left( \frac{\partial \beta_{\rm x_1}}
{\partial \theta_{\rm x_2}} + \frac{\partial \beta_{\rm x_2}}
{\partial \theta_{\rm x_1}} \right) .
\end{equation}
For the calculations shown
below, the spacing of the light rays on the
regular grid was taken to be 0.05 arcsec.

Unlike the circular lens, for which the shear is circularly--symmetric
as a function of radial distance from the lens center, the shear due
to an elliptical lens is a function of both the distance from the lens
center and the location of the source relative to the major and minor
axes of the mass distribution of the lens (as projected on the sky).
This effect is illustrated in Fig.~1, in which we plot the ratio
$\left< \gamma \right>_{\rm minor} / \left< \gamma \right>_{\rm major}$
as a function of angular scale, $\theta$. 
Here $\left< \gamma \right>_{\rm major}$ is the mean shear experienced
by sources closest to the major axis direction vectors of the lens and
$\left< \gamma \right>_{\rm minor}$ is the mean shear experienced by
sources closest to the minor axis direction vectors.  In the case of circularly
symmetric lenses this quantity will, of course, be unity 
on all scales.
In Fig.\ 1 we have computed
$\left< \gamma \right>_{\rm major}$ and 
$\left< \gamma \right>_{\rm minor}$
for sources located
within a polar angle of
$\pm N$ degrees relative to the axial direction vectors of the lens, for
the cases of
$N = 45^\circ$ (lefthand panel), and $N = 20^\circ$
(righthand panel). 
That is, in the righthand panel of Fig.~1
we consider only sources whose locations on the sky
are relatively close to the direction vectors defined by the major and
minor axes of the mass distribution while in the lefthand panel
we compute the mean
shear encountered by all sources.  Since all distance information
is contained within $\kappa_0$, the ratios shown in Fig.~1
are explicitly independent of the lens and source redshifts, as
well as the cosmology.
As expected, the degree of anisotropy in the shear pattern increases
with increasing lens ellipticity, and the closer the sources are to the 
axial direction vectors the larger is the level of the anisotropy.

\section{Prospects for Detection}

Provided the ellipticity of the light from a given candidate lens
galaxy is reasonably well--aligned with any flattening of its dark
matter halo, it is conceivable that one could investigate anisotropy
in the galaxy--galaxy lensing signal by first aligning the symmetry axes
of the lens galaxy images, then stacking the aligned images together 
in order to 
create one primary lens center about which the distortion of 
all the background galaxies 
could be measured (see, e.g., Natarajan \& Refregier 2000).  This will
be a reasonable procedure as long as the candidate lenses are in a 
fairly relaxed state (i.e., one would want to exclude
candidate lenses which have undergone a recent collision, for example).
The detectability of the anisotropy will, of course, depend not only
on the mean halo ellipticity, but also on the size of the data set
(i.e., to reduce the ``noise'' due the intrinsic shapes of the
background galaxies), the quality of the imaging data (i.e., the
accuracy with which image shapes can be determined), and the success with
which genuine foreground galaxies can be separated from genuine
background galaxies (i.e., the ability to discriminate lenses from sources).

The mean shear due to a (spherical)
singular isothermal
lens within a circular aperture of angular radius $\theta$ is:
\begin{equation}
\overline{\gamma}(\theta) = \frac{4 \pi}{\theta} \left( 
\frac{\sigma_v}{c} \right)^2 \left( \frac{D_{ds}}{D_s} \right) .
\end{equation}
So, for a fiducial lens with $\sigma_v = 155$ km/s, located at a redshift
of $z_d = 0.5$ and a fiducial source located at a redshift of 
$z_s = 1.0$, the mean shear will be 
$\overline{\gamma}(\theta) \sim 0.30/\theta''$.
(This result depends only weakly on the cosmology
through the ratio of $D_{ds}/D_s$.) 
At scale of $\theta = 25''$ this
is a small expected shear (of order 1\%), but it is certainly
measurable even with a modest--sized data set that has good imaging
quality.
The most statistically significant detection of galaxy--galaxy lensing to be
obtained via a direct average of the observed shapes of distant galaxies
is that reported by Fischer et al. (2000) from several nights of
Sloan Digital Sky Survey (SDSS) commissioning data. 
They have detected a net shear in the images of $\sim 1.5\times 10^6$
``faint'' galaxies ($r$ magnitudes in the range 18 to 22) due to
 $\sim 2.8\times 10^4$
``bright'' galaxies ($r$ magnitudes in the range 16 to 18) over an area
of nearly 225 sq.\ degrees and find a net 
shear of $\gamma \sim 0.005$ on an angular scale $\sim 30''$.  On
this angular scale, their detection is of order 6-$\sigma$. 

It is interesting to ask how large a data set would be required to 
obtain a significant detection of an anisotropy in the galaxy--galaxy lensing
signal in the manner we have computed here.  
For a given surface density
of lenses and sources, the signal to noise in
a measurement of $\left< \gamma \right> $ 
scales as the square root of the area of sky covered by the
imaging data (see, e.g.,  BBS
and Natarajan \& Refregier 2000).  If we compute
the mean shear experienced by sources within $\pm 45^\circ$ of the major axis
vectors of flattened halos, $\left< \gamma \right>_{\rm major}$,
and compare that to the shear experienced by
sources within $\pm 45^\circ$ of the minor axis vectors,
$\left< \gamma \right>_{\rm minor}$, the signal
to noise for each of these two quantities will be of order $1/\sqrt{2}$ times
the signal to noise for a measurement of the mean shear experienced by
all sources, as computed from a circular average about the lens centers.  That
is, the area over which 
$\left< \gamma \right>_{\rm major}$ or $\left< \gamma \right>_{\rm minor}$ 
is computed is only half of the area over 
which the circularly--averaged shear is
computed.  

As a rough attempt to calculate the detectability of an anisotropy in 
the galaxy--galaxy signal, we will restrict ourselves to halos that
have a mean flattening that is consistent with current observational
constraints: $\epsilon = 0.3$ (see, e.g., Sackett 1999 or Wright \&
Brainerd 2000).    In order to detect anisotropic galaxy--galaxy lensing
at a 4-$\sigma$ level from all sources within $\pm 45^\circ$ of
the axial direction vectors of a halo with $\epsilon = 0.3$, 
we need a signal to noise in the
value of $\left< \gamma \right>_{\rm minor} / 
\left< \gamma \right>_{\rm major}$ that is of order
$(0.76/0.06) \simeq 12.7$ (see, e.g., Fig.\ 1, lefthand panel).  
By straightforward error
propagation, we would then 
need a signal to noise in the (separate) measurements of
$\left< \gamma \right>_{\rm major}$ 
and $\left< \gamma \right>_{\rm minor}$ on the order of 18.  Assuming
an identical surface density of galaxies and identical noise properties
as the Fischer et al.\ data set, this would require an area of sky of order 
$\left[ 18 \times(\sqrt{2}/6) \right]^2 \simeq 18$
times that
used by Fischer et al.\ (2000).  That is, the amount of data required
would be equivalent to about 40\% of
the final SDSS data set ($\sim 10^4$ square degrees).
A somewhat larger
survey area (of order 26 times that of the Fischer et al.\ data
set) would be required to detect an anisotropy in 
$\left< \gamma \right>_{\rm minor}$
versus $\left< \gamma \right>_{\rm major}$ 
if one were to restrict the analysis to 
only those sources within $\pm 20^\circ$ of the axial direction vectors
of the lenses (e.g., Fig.\ 1, righthand panel).

One can, of course, improve the signal to noise by using a smaller area
and a data set with a completeness limit that is much greater than the
relatively shallow limit of the SDSS.  For example, BBS
claimed a 4-$\sigma$ detection of galaxy--galaxy lensing on scales
$\theta \ls 35''$ 
using a single, small field ($\sim 72$ sq.\ arcmin.) in which their
``bright'' galaxies had $r$ magnitudes in the range of
20 to 23 and their ``faint'' galaxies had $r$ magnitudes in the
range 23 to 24 (i.e., roughly 2 magnitudes fainter than the SDSS
data).  Applying the same
signal to noise calculation above to a survey with similar depth and
noise properties as the BBS data, an assumption of $\epsilon \sim 0.3$ 
then leads to an estimate of an area of order  
$\left[ 18 \times (\sqrt{2}/4) \right]^2 = 40.5$ times that of the 
BBS data set (i.e., only about 0.8 square degrees) 
being necessary in order to
obtain a significant detection of an anisotropy in the galaxy--galaxy lensing
signal using sources within $\pm 45^\circ$ of the axial direction
vectors of the lenses.  A survey of area 1.8 square degrees would be
necessary to obtain a 4-$\sigma$ detection of
anisotropy in the signal using sources within
$\pm 20^\circ$ of the axial direction vectors of the lenses.  Such relatively
small areas can now be obtained easily with wide-field CCD mosaic cameras
and, therefore, such an investigation is well within the reach of
current technology. 

There are, however, some points to note when considering
the above estimation of the detectability of anisotropic galaxy--galaxy
lensing.  The first, and most obvious, is that both 
Fischer et al.\ (2000) and BBS separated
candidate lenses from candidate sources on the basis of apparent 
magnitude alone.
That is, since galaxies have a broad distribution in redshift,
lens--source separation performed solely 
on the basis of apparent magnitude will
be inefficient and some of the candidate lenses are, therefore,
located at greater distances from the observer
than are some of the candidate
sources.  This will, necessarily, manifest as a source of ``noise''
in the analysis and will degrade the lensing signal.
If lens--source separation on the basis of photometric or spectroscopic
redshifts were to be performed, for example, the size of the data set that
would be required to detect anisotropic galaxy--galaxy lensing would
be reduced significantly from our calculation above.

Another point to note is that our estimation of the detectability of
the signal is based upon an assumption of single deflections.  That
is, in computing the theoretical values of 
$\left< \gamma \right>_{\rm minor} / \left< \gamma \right>_{\rm major}$
we have not accounted for the fact that distant galaxies may be
weakly lensed at comparable levels by two or more foreground galaxies.
The number of multiple deflections that need to be considered depends
upon the selection function used to discriminate between candidate
lenses and sources, but in the case of the magnitude--selected samples
of BBS it was estimated that  1/3
of the faint galaxy sample would have been lensed at a comparable level
by 2 foreground galaxies, and that another 1/3 of the faint galaxy
sample would have been lensed at a comparable level by 3 or more
foreground galaxies.  Therefore, a somewhat more reliable estimate of the
detectability of anisotropic galaxy--galaxy lensing will depend upon
a more detailed analysis of the problem than we have presented here.
In particular, simulations of galaxy--galaxy lensing which
match observational constraints as closely as possible (i.e., the faint
galaxy number counts, the
redshift distribution of galaxies as a function of magnitude, the
luminosity function of galaxies, the range of reasonable halo shapes, and
the noise properties of the data) will ultimately
be required.  We are in the process of completing such an analysis
and will present the results soon (Wright \& Brainerd 2000). 

Lastly, the proposed
``stacking'' of the foreground galaxy images will, necessarily,
be noisy at some level and this will contribute to a degradation of the
anisotropic lensing signal.  In particular, the
effects of seeing, pixellation, and sky noise will
all contribute an error to the observed position angle of the image of
a foreground galaxy.  In addition, galaxy--galaxy
lensing of the selected foreground
population due to nearby galaxies along the line of sight
will contribute a small but
non-zero error.  That is, even in the limit of ``perfect'' imaging 
data, galaxy--galaxy lensing of the foreground population itself
will cause the observed position angle of the image of a foreground
galaxy to differ slightly from the true position angle of the galaxy.
However, this effect is expected to be small compared to the error in the
position angle that is induced by the imaging process.  For example,
the Monte Carlo simulations performed by BBS show that for 80\% of 
galaxies, the position angle of the lensed image differs from that of
the unlensed image by less than $5^\circ$.

In order to estimate the degree to which noise in the determination
of the position angles of the foreground galaxies results in a degradation
of the anisotropic lensing signal, we repeat the calculations in \S2 but
with the difference that we induce an error in the observed orientation
of the axial direction vectors by randomly rotating them
away from their ``true'' location.   That is, the shear experienced by
the source galaxies is correctly computed as being due to flattened lenses
with position angles of $0^\circ$.  Then, the orientation of the axial direction
vectors is randomly rotated by an amount $\phi$
and the mean shears, $\left< \gamma 
\right>_{\rm major}$ and $\left< \gamma \right>_{\rm minor}$, are computed
using sources within $\pm N^\circ$ of these rotated direction vectors
(i.e., the ``observed'' direction vectors).
For simplicity of calculation, the values of $\phi$ were drawn from Gaussian
distributions with zero mean and standard deviations of $\sigma$, where
$\sigma$ ranged from $5^\circ$ to $45^\circ$.  Computing the anisotropy
in the lensing signal relative to the rotated axes will necessarily result in
a value of 
$\left< \gamma \right>_{\rm minor} / \left< \gamma \right>_{\rm major}$
that is closer to unity than are the values plotted in Fig.\ 1.
That is, the anisotropic lensing signal will be ``degraded'' by some amount,
which will clearly be dependent upon the typical error in the position angles
of the lens galaxies, $\phi$.

By symmetry, 
the ratio of the mean shears is independent
of angular scale (e.g., Fig.~1) for any randomly chosen value of
$\phi$ and we therefore do not plot the angular dependence of the
anisotropic lensing signal as measured relative to the rotated axes.
For each value of lens ellipticity, $\epsilon$,
and standard deviation in the position angle, $\sigma$, we compute the
ratio,
\begin{equation}
\zeta = 
\left[1- {\left< \gamma \right>_{\rm minor} \over 
\left< \gamma \right>_{\rm major}} \right]_{\rm obs} 
\left[1- {\left< \gamma \right>_{\rm minor} \over 
\left< \gamma \right>_{\rm major}} \right]_{\rm true}^{-1} 
\end{equation}
where the numerator is the mean value of the anisotropic lensing
signal that was obtained from 5000 independent
rotations of the axial direction vectors
and the denominator is obtained from the
values plotted in Fig.\ 1.  That is, here we compare the relative deviations
of the lensing signals from values of
unity (the null case in which all halos are
round in projection).
The results are
shown in Fig.\ 2, where the lefthand panel shows the degradation in the
lensing signal for sources within $\pm 45^\circ$ of the rotated axial direction
vectors and the righthand panel shows the same, but for sources within
$\pm 20^\circ$ of the rotated
axial direction vectors.  The different point types
refer to lenses of different ellipticities and correspond to the 
point types used in Fig.\ 1.
It is clear from Fig.\ 2 that the anisotropic lensing signal is degraded
due to the noise associated with
stacking the images of the foreground lenses, but provided
the typical 1-$\sigma$ error in the position angle of the lenses is less than
$15^\circ$, more than 90\% of the true signal is recovered in the
mean.  The more
accurate the alignment of the foreground galaxies, the more signal that
will, necessarily, be recovered.  Therefore,
even with somewhat noisy data, anisotropies in the galaxy--galaxy lensing
signal should still be detectable.

\section{Summary}

We have investigated anisotropies in galaxy--galaxy lensing by comparing
the mean shear experienced by sources nearby to the major axis of an
elliptical lens to that experienced by sources nearby to the minor axis.
We have shown that for realistic halo flattening ($\epsilon \sim 0.3$),
the level of anisotropy in the lensing signal is small but should be
detectable in large ground-based imaging surveys.  Our calculation
of the size of the data set required for detection of this effect includes
the actual observational
error estimates in the detection of circularly-averaged galaxy-galaxy
lensing from two previous investigations of galaxy--galaxy
lensing: Fischer et al.\ (2000) and BBS.  In 
addition, we have shown that although the level of the detected anisotropy is
degraded due to noise in the process of aligning the symmetry axes of the
foreground lens galaxies prior
to stacking their images, the effect should be small provided the true position
angles of the lens galaxies are known to within a typical 1-$\sigma$
error of order
$15^\circ$.

Galaxy--galaxy lensing is the only technique which at present has the
potential to constrain the projected shapes of the dark matter galaxy halos 
on average throughout the universe.  
While it cannot provide a strong constraint on the shape of
any one particular halo, 
the method is applicable to the entire galaxy population and should
yield a strong constraint on the mean projected shape of halos.  Given
that there are currently very few observational constraints on the 
shapes of dark matter halos, the wealth of information on both
the details of galaxy formation and the nature of the dark matter that
such constraints will provide, and the apparent detectability of
the lensing signal for moderately flattened halos,
 observational investigations of 
anisotropic galaxy--galaxy lensing with high-quality imaging data
certainly appear to be justified at this time.

\section*{Acknowledgments}

Support under NSF contract AST-9616968 
and a generous allocation of resources at Boston University's
Scientific Computing and Visualization Center are gratefully acknowledged.

\clearpage

\begin{figure}
\plotfiddle{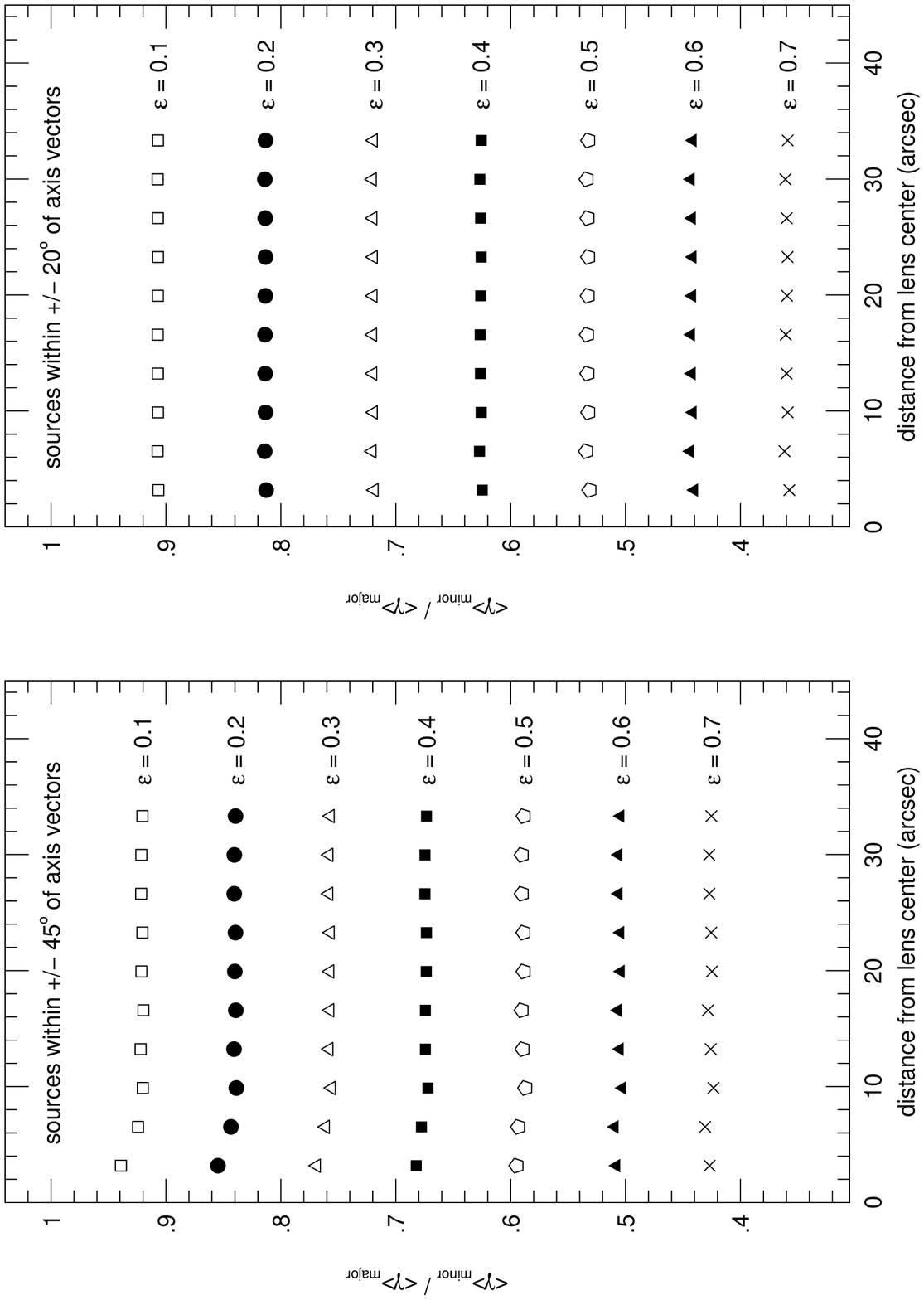}{5.0in}{0}{60.0}{60.0}{-250}{50}
\vspace{-1.0in}
\caption{
Ratio of the mean shear experienced by sources closest to the
minor axis of an elliptical lens to that experienced by sources 
closest to the major axis of the lens.   Lefthand panel: all sources that would
be found within $\pm 45^\circ$ of the axial direction vectors are included
in the calculation.
Righthand panel: all sources that would
be found within $\pm 20^\circ$ of the axial direction vectors are included
in the calculation.
}
\end{figure}

\clearpage
\begin{figure}
\plotfiddle{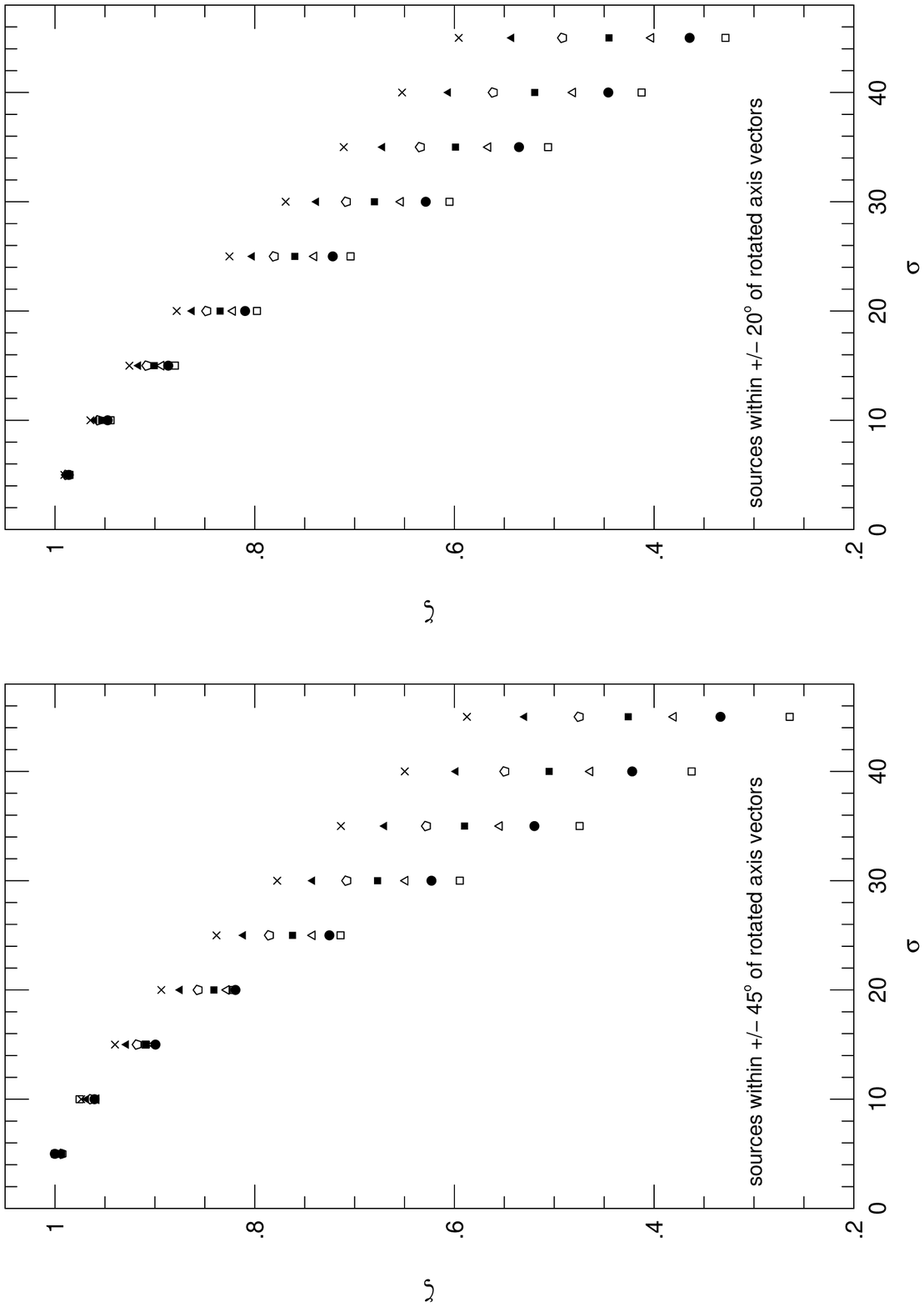}{5.0in}{0}{60.0}{60.0}{-250}{50}
\vspace{-1.0in}
\caption{Comparison of
the anisotropic lensing signal for lenses whose image position
angle has been randomly rotated away from the true position angle and
the lensing signal obtained without errors in
the position angle (e.g., equation 6 above).
The different point types indicate lenses with different ellipticities
and correspond to the point types used in Fig.\ 1.  For each value of 
$\sigma$, the axial direction vectors of the
lens were randomly rotated 5000 times by an amount
$\phi$, where $\phi$ was drawn from a Gaussian distribution with zero
mean and a standard deviation of $\sigma$.  The lefthand panel shows the
result for sources within $\pm 45^\circ$ of the rotated axial direction
vectors; the righthand panel shows the same, but for sources within
$\pm 20^\circ$.
}
\end{figure}


\clearpage
\begin{references}
\reference{bbs} Brainerd, T.G., Blandford, R. D. \&
Smail, I., 1996, \apj, 466, 623  (BBS)
\reference{} Dell'Antonio, I. P. \& Tyson, J. A. 1996, \apj, 473, L17
\reference{} Dubinski, J. \& Carlberg, R. G. 1991, \apj, 378, 496
\reference{} Ebbels, T. 1998, PhD Thesis, University of Cambridge
\reference{} Fich, M. \& Tremaine, S. 1991, \araa, 29, 409
\reference{} Fischer, P. et al. (the SDSS Collaboration), 2000, \aj, submitted
(astro-ph/9912119)
\reference{} Griffiths, R. E., Casertano, S., Im, M., \& Ratnatunga, K. U.
1996, \mnras, 282, P1159
\reference{} Hudson, M. J., Gwyn, S. D. J., Dahle, H., \& Kaiser, N. 1998,
\apj, 503, 531
\reference{} Kochanek, C. S., Keeton, C. R., \& McLeod, B. A., 2000, 
astro-ph/0006166
\reference{} Moore, B., Gelato, S., Jenkins, A., Pearce, F. R., \&
Quilis, V. 2000, astro-ph/0002308
\reference{priya} Natarajan, P., Kneib, J.-P., Smail, I., \& Ellis, R. S, 
1998, \apj, 499, 600
\reference{} Natarajan, P. \& Refregier, A. 2000, astro-ph/0003344
\reference{} Sackett, P. D. 1999, in {\it Galaxy Dynamics}, ASP conference
series vol.\ 182, eds.\ D. R. Merritt, M. Valluri, \& J. A. Sellwood, 393
\reference{sw91} Schneider, P. \& Weiss, A. 1991, A\&A, 247, 269 (SW91)
\reference{} Warren, M. S., Quinn, P. J., Salmon, J. K., \& Zurek, W. H.
1992, \apj, 399, 405
\reference{} Wright, C. O. \& Brainerd, T. G. 2000, in preparation
\reference{} Zaritsky, D. \& White, S. D. M. 1994, \apj, 435, 599
\reference{dennis2} Zaritsky, D., Smith, R., Frenk, C., \& White, S. D. M.,
1997, \apj, 478, 39

\end{references}
\end{document}